\documentclass[pdflatex,sn-basic]{sn-jnl}

\usepackage{graphicx}%
\usepackage{multirow}%
\usepackage{amsmath,amssymb,amsfonts}%
\usepackage{amsthm}%
\usepackage{mathrsfs}%
\usepackage[title]{appendix}%
\usepackage{xcolor}%
\usepackage{textcomp}%
\usepackage{manyfoot}%
\usepackage{booktabs}%
\usepackage{algorithm}%
\usepackage{algorithmicx}%
\usepackage{algpseudocode}%
\usepackage{listings}%
\usepackage{caption}%
\usepackage{subcaption}
\usepackage{graphicx}

\theoremstyle{thmstyleone}%
\theoremstyle{thmstyletwo}%

\theoremstyle{thmstylethree}%

\begin{document}

\title[Article Title]{Contextual Factors Behind Income Inequality: \\A Predictive Analysis}


\author[1]{\fnm{Arda} \sur{Aktas}}\email{a.aktas@vu.nl}\equalcont{These authors contributed equally to this work.}

\author*[1]{\fnm{Miguel} \sur{{Poblete-Cazenave}}}\email{m.a.poblete.cazenave@vu.nl}

\author[2,3]{\fnm{Narasimha} \sur{Rao}}\email{narasimha.rao@yale.edu}
\equalcont{These authors contributed equally to this work.}

\affil*[1]{\small\orgdiv{Institute for Environmental Studies (IVM)}, \orgname{Vrije Universiteit Amsterdam}, \orgaddress{\street{De Boelelaan 1111}, \city{Amsterdam}, \postcode{1081 HV}, \country{The Netherlands}}}

\affil[2]{\small\orgdiv{Yale School of the Environment}, \orgname{Yale University}, \orgaddress{\street{195 Prospect Street}, \city{New Haven}, \postcode{CT 06511}, \state{Connecticut}, \country{USA}}}

\affil[3]{\small\orgdiv{Energy, Climate, and Environment Program}, \orgname{International Institute for Applied Systems Analysis (IIASA)}, \orgaddress{\street{Schlossplatz 1}, \city{Laxenburg}, \postcode{A-2361}, \state{Lower Austria}, \country{Austria}}}

\abstract{Income inequality remains a central concern in economics and public policy, yet conclusions about its predictors often depend on both the factors considered and the measure of inequality used. This paper examines the predictive importance of a broad set of indicators covering structural factors, capital structure, financial-sector development, political institutions, regional classification, human capital, globalization, and technology. Using data for 88 countries from 1970 to 2017 and six inequality measures from the World Income Inequality Database (WIID), we apply a machine learning approach to assess the relative predictive importance of 35 candidate indicators across alternative income inequality measures.

The analysis proceeds in three stages: a full specification including all candidate indicators, a category-representative specification with one common indicator from each conceptual category, and an inequality-specific specification allowing the most predictive indicator within each category to vary across inequality measures.

Regional classification is the dominant predictor across all specifications and inequality measures, although its broad nature prevents identification of the specific regional characteristics underlying this result. Political and technology indicators provide the most consistent additional predictive information, whereas the importance of human capital, financial development, and globalization varies across inequality measures. Technology is particularly informative for top-income concentration, while returns to capital are more relevant for measures capturing gaps between the upper and lower parts of the income distribution.
Despite these differences, the category-representative specification retains most of the full model's predictive information, showing that cross-country inequality can be analysed using a more parsimonious set of indicators without losing much predictive power.} 

\keywords{income inequality, XGBoost, Machine learning, political institutions, regional classification, cross-country analysis}



\maketitle

\section{Introduction}\label{sec:intro}
Income inequality has been a persistent economic and policy concern because income disparities affect people's access to goods and services, including those necessary to maintain an adequate standard of living. For a few decades in the mid-20th century, income inequality appeared to decline in many countries while average incomes rose, creating the impression that sustained economic growth alone might gradually reduce the problem. Since the 1980s, however, this pattern has reversed, with income inequality increasing across many parts of the world, including high-income countries with sufficient resources to support broadly shared prosperity.

These developments have brought the longstanding question of what drives inequality back into focus. A large body of literature has examined this issue, with notable contributions dating back to the seminal work of Kuznets \citep{Kuznets1955}, which relates economic growth to income inequality by arguing that inequality increases during the early stages of economic development and begins to decline after a certain level of income is reached, producing the well-known inverted U-shaped Kuznets curve. However, this proposition has become increasingly difficult to reconcile with observed trends, particularly in recent decades, as income inequality has risen rapidly in some of the world’s largest economies. During this period, the structure of production has changed considerably, along with major shifts in labor markets, financial systems, political institutions, technology, and globalization. Therefore, income growth alone is no longer sufficient to explain whether inequality rises or falls, leaving room for a range of other factors that may help explain different parts of the puzzle.

Among the various studies proposing different drivers of inequality in the recent literature, the work of Piketty and co-authors \citep{Piketty2001,Piketty2003,Piketty2014} has attracted particular attention. Their central argument is that if the rate of return on capital exceeds the rate of economic growth, incentives for capital accumulation will tend to drive inequality upward. While this view has been subject to some critique \citep{Krusell2015,Homburg2015}, there appears to be sufficient theoretical and empirical evidence supporting some of its key elements. However, while the argument largely explains rising wealth inequality, it falls short of explaining the observed increase in income inequality and therefore remains far from providing a complete picture of inequality \citep{Acemoglu2015,Goes2025}.

A broader view considers a range of economic, institutional, and structural factors that affect income generation and distribution. Starting with the labor force,  one of the most important factors contributing to inequality is variation in human capital. Workers differ in their productivity, and these differences are reflected in their wages, an idea that dates back to Adam Smith \citep{Spengler1977}. Because individual productivity cannot be directly observed, other measures are often used to assess potential productivity differences \citep{Folloni2010,Abraham2022}. Nevertheless, the widening gap between wage growth and growth in GDP per capita points to factors beyond human capital as potential drivers of inequality \citep{Brynjolfsson2014,Berlingieri2024}.

Technology appears to be one such factor, either by increasing productivity or by reducing the role of labor in production \citep{Caselli1999,Acemoglu2011}. Globalization may also contribute to inequality by expanding the cross-border movement of goods, capital, and economic activity and increasing firms’ ability to organize production across countries. Because capital is more mobile internationally than labor, firms may relocate production to countries with lower labor costs when workers there can perform same tasks at comparable levels of productivity \citep{Helpman2010,Autor2013}. Government and political institutions represent another important factor. Institutional frameworks can influence both development and inequality by affecting access to economic opportunities and determining how the gains from economic activity are distributed, as explored extensively by Acemoglu, Robinson, and co-authors \citep{Acemoglu2005,Acemoglu2012}. These institutional arrangements, however, are also embedded in broader historical and regional contexts. Countries within the same region often share political, legal, and institutional legacies that may contribute to similar patterns of inequality \citep{Andreescu2025}. Finally, the financial sector may also play an important role. One possible channel links back to Piketty’s argument: if capital owners receive higher returns in financial markets than in productive activities, capital will increasingly flow toward the financial sector, contributing to greater inequality \citep{Greenwood1990,deHaan2017}.

In sum, many of the factors associated with inequality are broad conceptual categories rather than single, clearly defined mechanisms. Human capital, for example, may be represented by average educational attainment, the distribution of education, health, or a composite human capital index, while political institutions may differ in their electoral, liberal, participatory, deliberative, or egalitarian dimensions. Similarly, technology, globalization, and financial development each encompass several distinct features that need not be equally informative about inequality. The conclusions reached may therefore depend on both the indicator used to represent each broader category and the measure of inequality being examined \citep{Acemoglu2015}. Consequently, identifying which indicators are most informative under alternative inequality measures is important for obtaining a clearer picture of inequality dynamics.

In this study, we use modern predictive techniques to identify which indicators within these broader conceptual categories are most informative for predicting income inequality in the following year across alternative inequality measures. Using a broad cross-country panel dataset and a diverse set of indicators capturing the main factors discussed in the literature, we employ a machine-learning approach to evaluate their relative predictive importance. This approach is particularly suited to our setting because it can accommodate a large set of potentially correlated predictors, as well as nonlinear relationships and interactions, without imposing a predetermined parametric structure. Rather than focusing on causal identification or the estimation of structural coefficients, our objective is to assess which indicators consistently contain information about future inequality outcomes and whether their relevance varies across different parts of the income distribution.

Our contribution is threefold. First, we simultaneously evaluate a broad range of factors emphasized in the literature on income inequality, including political institutions, human capital, globalization, financial development, and technology. Second, we examine six commonly used measures of income inequality, allowing us to assess whether the factors associated with overall inequality differ from those associated with top-income concentration and relative income gaps. By considering several measures, the study provides a broader view of the factors associated with different parts of the income distribution than analyses based on a single measure. Third, by adopting a predictive rather than causal perspective, we provide complementary evidence on which contextual factors consistently contain information about inequality across countries and which are more relevant to particular measures of inequality.


\section{Data and Variables}\label{sec:data}
We combine information from multiple international datasets and compile a broad set of indicators capturing factors that have been associated with income inequality in the literature.

The indicators are described below and summarized in Table \ref{tab:indicators}, where they are organized into thematic categories reflecting the main conceptual channels discussed in the literature. From the set of potential indicators, we include only those with observations covering at least 60\% of the country–year sample. This threshold was chosen to retain sufficient data coverage while preserving representation across the different conceptual categories.

\subsection{Income Inequality Measures}
Income inequality can be measured in several ways, and because different measures capture different parts of the income distribution, empirical conclusions may vary depending on the measure used. This is reflected, for example, in the contrasting interpretations of inequality associated with Piketty \citep{Piketty2014} and the critique by Acemoglu and Robinson \citep{Acemoglu2015}

To account for these differences, we employ six alternative measures of income inequality, all obtained from WIID and widely used in the empirical literature. They are defined below, together with references to seminal or widely cited studies employing each measure:

\begin{itemize}
\item[-] The Gini coefficient  \citep{Gini1936,Sen1973}
\item[-] The income share held by the top 10\% of the population \citep{Atkinson2007,Leigh2007}
\item[-] The income share held by the top 1\% of the population \citep{Piketty2003,Acemoglu2015}
\item[-] The Palma Ratio: the share held by the top 10\% divided by the share held by the bottom 40\% \citep{Palma2011,Cobham2014}
\item[-] The S80/S20 ratio (also called the 20:20 ratio): the share held by the top 20\% divided by the share held by the bottom 20\% \citep{UNDP1999,Forbes2000}
\item[-] The P90$/$P10 ratio:income share at the 90th percentile divided by income share at the 10th percentile of the income distribution \citep{Piketty2001,Autor2008}
\end{itemize}

The measures are drawn from the World Income Inequality Database (WIID) \citep{WIID2023}, compiled by the United Nations University World Institute for Development Economics Research (UNU-WIDER). WIID is one of the most comprehensive global inequality databases, providing harmonized Gini coefficients and income share statistics derived from household surveys, tax records, and national statistical sources. 

\subsection{Capital Structure}
We include two variables calculated using data from the Penn World Tables (PWT) \citep{Feenstra2015} to represent Piketty's view of inequality as being driven by the structure of capital: the capital-to-labor ratio  ($KY\_L$), and the ratio of returns to capital to GDP growth ($r\_g$), which shows whether returns to capital exceed the rate of economic growth. 

\subsection{Financial Sector}
To measure financial-sector development, we use a set of variables from the World Bank’s Global Financial Development Database (GFDD) \citep{Beck2000,WB2019}. These include private credit to GDP ($di01$ and $di12$), deposit money banks’ assets to GDP ($di02$), liquid liabilities relative to GDP ($di05$), financial system deposits to GDP ($di08$ and $oi02$), the ratio of bank credit to bank deposits ($si04$), and total liquid liabilities ($oi07$). 

\subsection{Political Institutions}
To assess which aspects of political institutions are most informative for predicting inequality, we use a set of indicators from the V-Dem database, including the electoral democracy index ($v2x\_polyarchy$), the liberal component index ($v2x\_liberal$), the participatory component index ($v2x\_partip$), the deliberative component index ($v2xdl\_delib$), and the egalitarian component index ($v2x\_egal$).

These indicators provide alternative characterizations of political institutions, focusing on the quality of elections and political competition ($v2x\_polyarchy$), the protection of minority rights, civil liberties, and constraints on government power ($v2x\_liberal$), citizen participation in political life within and beyond elections ($v2x\_partip$), the role that public reasoning toward the common good plays in political decision-making ($v2xdl\_delib$), and equality in political rights, resource distribution, and access to power across social groups ($v2x\_egal$).

\subsection{Regional Aggregation}
We use several regional classifications to group countries that share broad historical, political, and geographic characteristics that may affect development and income inequality in similar ways within each region. These include the 23 World Bank subregions ($region\_wb$) and the 18 United Nations subregions ($un\_subregion$), both obtained from the \textit{R} \textit{countrycode} package \citep{countrycode}.

From the V-Dem dataset, we also include a politico-geographic region classification ($e\_regionpol$), originally obtained from the Quality of Government Standard Dataset (QoG) \citep{Teorell2013}. This classification groups countries into 10 regions based on a combination of geographical proximity and regional boundaries defined by area specialists’ understanding of democratization.

We include two additional aggregation variables as controls, to test whether the predictive importance of regional aggregation reflects more specific historical or economic similarities. From the QoG dataset, we use $ht\_colonial$, which indicates whether a country was formerly colonized and by which major colonial power. We also add the World Bank income-group classification ($incomegroup$) to test whether the predictive importance of the regional classifications for income inequality simply comes from similarities in countries’ income levels.  

\subsection{Human Capital}
We use the human capital index from the Penn World Table ($hc$), which combines years of schooling with estimated returns to education, together with three indicators from the V-Dem database: average years of education among individuals aged 15 and older ($e\_peaveduc$), an index of educational inequality among the adult population ($e\_peedgini$), and life expectancy ($e\_pelifeex$). For some countries, the latter two indicators were complemented with data from the Wittgenstein Centre Human Capital Data Explorer (WCDE) \citep{KC2024}. Although the WCDE provides more comprehensive country coverage than the V-Dem database, its data are available only at five-year intervals and therefore require interpolation.

\subsection{Globalization}
We obtain globalization indicators from the KOF Globalization Index \citep{Dreher2006,Gygli2019}, which covers economic, trade, and financial globalization and distinguishes between actual cross-border flows (de facto globalization) and the institutional framework governing international integration (de jure globalization).

\subsection{Technology}
We consider several indicators related to technological sophistication, productivity, and the contribution of technology to economic activity.

Economic Complexity Index (ECI), obtained from The Observatory of Economic Complexity (OEC) \citep{Simoes2011,ECI}, shows a country’s stock of productive knowledge based on composition and sophistication of its export basket. It increases when a country exports many different products, particularly those that few other countries can produce. Therefore, higher values indicate a more advanced and knowledge-intensive productive system \citep{Hidalgo2009}. 

We also use two productivity-based measures obtained from the Penn World Table (PWT), version 11.0: standard Total Factor Productivity \textit{ctfp} and welfare-adjusted TFP\textit{cwtfp}. The first one measures how efficiently capital and labor are used to produce output and is therefore commonly interpreted as an indicator of technological efficiency. The second also incorporates terms-of-trade effects and better reflects differences in welfare and purchasing power across countries.

Because there is no consensus on how to calculate TFP, we also include an alternative measure, $tfp\_alt$, calculated using the growth of total factor productivity obtained from the Total Economy Database \citep{TheConferenceBoardTotalEconomyDatabase2025}. Unlike the TFP measure from the PWT that is calculated using the growth of the capital stock, this measure is obtained from the growth of capital services. From the same dataset, we also use two additional measures related to digitalization of the economy: $kgdpgric$, which shows the contribution of capital services provided by information and communication technology (ICT) assets to real GDP growth, and $kgdpgrnic$, which shows the contribution of non-ICT capital services to real GDP growth.

\subsection{Other Structural Factors}
We include a final set of indicators related to the broader economic, demographic, and state of unrest conditions of each country. Among these, gross domestic product (GDP) and total population, which represent overall economic size, country scale, and market size, are obtained from the World Income Inequality Database (WIID).

Urbanization rates ($e\_miurbani$), which measure the share of the population living in urban areas and are considered an indicator of the spatial distribution of population and economic activity, are obtained from the Varieties of Democracy (V-Dem) dataset \citep{VDEM2025} and the World Development Indicators \citep{WDI}. 

Finally, we add indicators of international armed conflict ($e\_miinteco$) and internal armed conflict ($e\_miinterc$), originally from the Uppsala Conflict Data Program (UCDP) \citep{Sundberg2013,Davies2025}  and available in the V-Dem dataset only up to 2000, considering they may affect income inequality by disrupting economic activity and labor markets, destroying assets and infrastructure, displacing populations, and changing the allocation of public resources.

\begin{table}[h]
\caption{Set of Explanatory Variables*}\label{tab:indicators}%
\fontsize{5}{6}\selectfont
\begin{tabular}{@{}lll@{}}
\toprule
Variable & Description & Source \\
\midrule
\multicolumn{3}{l}{Structural Factors}  \\
\midrule
$gdp$ & Gross Domestic Product (GDP) & WIID \\
$population$ & Total Population & WIID \\
$e\_miurbani$ & Urbanization Rate: share of the population living in urban areas & V-Dem \& WDI \\
$e\_miinteco$ & Country's participation in an international armed conflict & V-Dem \& UCDP \\
$e\_miinterc$ & Country's Participation in an internal armed conflict & V-Dem \& UCDP \\
\midrule
\multicolumn{3}{l}{Capital Structure}  \\
\midrule
$KY\_L$ & Capital to Labor ratio$^\ddagger$ & PWT \\
$r\_g$ & Returns to capital over GDP growth$^\ddagger$ & PWT \\
\midrule
\multicolumn{3}{l}{Financial Sector}  \\
\midrule
$di01$ & Private credit by deposit money banks to GDP & GFDD \\
$di02$ & Deposit money banks' assets to GDP & GFDD \\
$di05$ & Liquid liabilities to GDP & GFDD \\
$di08$ & Financial system deposits to GDP & GFDD \\
$di12$ & Private credit by deposit money banks and other financial institutions to GDP & GFDD \\
$si04$ & Bank credit to bank deposits & GFDD \\
$oi02$ & Bank deposits to GDP& GFDD \\
$oi07$ & Liquid liabilities in millions USD (2010 constant) & GFDD \\
\midrule
\multicolumn{3}{l}{Political Institutions}  \\
\midrule
$v2x\_polyarchy$ & To what extent is the ideal of electoral democracy in its fullest sense achieved & V-Dem \\
$v2x\_liberal$ & To what extent is the liberal principle of democracy achieved & V-Dem \\
$v2x\_partip$ & To what extent is the participatory principle achieved & V-Dem \\
$v2xdl\_delib$ & To what extent is the deliberative principle of democracy achieved & V-Dem \\
$v2x\_egal$ & To what extent is the egalitarian principle achieved & V-Dem \\
\midrule
\multicolumn{3}{l}{Regional Aggregation}  \\
\midrule
$region\_wb$ & Regions as used to be in the World Bank Development Indicators & countrycode \\
$un\_subregion$ & United Nations sub-region & countrycode \\
$e\_regionpol$ & Politico-geographic classification of world regions & V-Dem \\
$ht\_colonial$ & Tenfold classification of former colonial ruler of the country, focused on 'Western overseas' colonialism & QoG \\
$incomegroup$ & World Bank classification of countries by income level & WIID \\
\midrule
\multicolumn{3}{l}{Human Capital}  \\
\midrule
$e\_peaveduc$ & Average years of education among individuals aged 15 and older& V-Dem \& WCDE$^\ddagger$ \\
$e\_peedgini$ & How unequal is the level of education of the population older than 15  & V-Dem \& WCDE$^\ddagger$ \\
$e\_pelifeex$ & Life Expectancy & V-Dem \& WCDE$^\ddagger$ \\
$hc$ & Human capital index, based on years of schooling and returns to education & PWT \\
\midrule
\multicolumn{3}{l}{Globalization}  \\
\midrule
$KOF\_econ$ & KOF Economic Globalization Index & KOF \\
$KOF\_defacto$ & De facto component of the KOF Economic Globalization Index & KOF \\
$KOF\_dejure$ & De jure component of the KOF Economic Globalization Index & KOF \\
$KOF\_trade$ & KOF Trade Globalization Index & KOF \\
$KOF\_trade\_df$ & De facto component of the KOF Trade Globalization Index & KOF \\
$KOF\_trade\_dj$ & De jure component of the KOF Trade Globalization Index & KOF \\
$KOF\_finance$ & KOF Financial Globalization Index & KOF \\
$KOF\_finance\_df$ & De facto component of the KOF Financial Globalization Index & KOF \\
$KOF\_finance\_dj$ & De jure component of the KOF Financial Globalization Index & KOF \\
\midrule
\multicolumn{3}{l}{Technology}  \\
\midrule
$eci$ & Economic Complexity Index, based on the diversity and sophistication of a country’s export basket & ECI \\
$ctfp$ & Total Factor Productivity (TFP) level at current purchasing power parity (PPT) (USA=1) & PWT \\
$cwtfp$ & Welfare-relevant TFP levels at current PPPs (USA=1) & PWT \\
$tfp\_alt$ & Alternative TFP measure calculated using total factor productivity growth from TED $^\ddagger$ & TED \\
$kgdpgric$ & Contribution of capital services provided by ICT Assets to real GDP growth & TED \\
$kgdpgrnic$ & Contribution of capital services provided by non-ICT Assets to real GDP growth & TED \\
\botrule
\end{tabular}
\footnotetext{Sources: World Income Inequality Database (WIID), Penn World Tables (PWT) v11.0, KOF Globalization Index (KOF), World Bank Global Financial Development Database (GFDD), World Development Indicators (WDI), Wittgenstein Centre
Human Capital Data Explorer (WCDE), Total Economy Database™ (TED), Uppsala Conflict Data Program (UCDP), Varieties of Democracy (V-Dem) dataset, Quality of Government (QoG) Standard Dataset, and the $R$ countrycode package. Variables are retained if coverage exceeds 60\% of the country-year sample.}
\footnotetext[*]{The variables are grouped into thematic categories 
reflecting key theoretical channels through which inequality may be affected.}
\footnotetext[\ddagger]{Variables calculated by the authors using the corresponding data.}
\end{table}


\section{Methodology}\label{sec2}
To assess the predictive importance of the different indicators presented in Table \ref{tab:indicators} within each conceptual category across alternative measures of inequality, we employ a machine learning approach based on the \textit{XGBoost} model \citep{Chen2016}. \textit{XGBoost} is a gradient boosting framework that is well suited to settings with multiple, potentially correlated predictors and can accommodate nonlinear relationships and interaction effects without requiring these relationships to be specified in advance. This flexibility allows us to assess the predictive relevance of the indicators without imposing a predetermined functional form.

The analysis is conducted using a country--year sample restricted to observations with complete information on the selected indicators, covering 88 countries from 1970 to 2017. The dependent variable is defined as next-year inequality, measured using six alternative indicators (the Gini coefficient, the top 10\% income share, the top 1\% income share, the Palma ratio, the 20:20 ratio, and the P90\/P10 ratio), allowing us to evaluate one-step-ahead predictive performance. All models are estimated using a random 70\% training sample over 50 bootstrap iterations to ensure the robustness of the results. The same training–-hold-out split is used across specifications within each bootstrap iteration to ensure comparability. 

We implement three complementary model specifications, which progressively refine the selection of predictors while preserving the conceptual category structure.

First, we estimate a baseline model that includes the full set of 35 candidate indicators simultaneously. This unrestricted specification allows us to evaluate and rank the relative predictive importance of each indicator for each inequality measure separately.

Second, to identify the indicator that has consistently more predictive power within each theoretical category across the alternative inequality measures, we construct a reduced specification based on cross-measure importance aggregation. More specifically, feature importance is scaled within each model and then averaged across the six inequality measures. Based on these average importance scores, we select one indicator from each of the following six conceptual categories: financial sector, human capital, globalization, political institutions, regional aggregation, and technology. In addition, we retain the most predictive baseline control variables. This specification preserves representation across the main theoretical categories while selecting a common set of predictors for all inequality measures.

Finally, we allow the optimal predictor within each of the six conceptual categories to differ across inequality measures. For each inequality measure, we perform a constrained model search in which we evaluate all possible combinations containing exactly one candidate indicator from each conceptual category, together with the retained baseline controls. For each combination, we estimate an \textit{XGBoost} model on the training sample and select the specification with the lowest combined mean squared prediction error across the training and hold-out samples in each bootstrap iteration. The procedure is repeated over 50 bootstrap iterations with different samples to assess the stability of the selected specifications and their predictive performance. This approach identifies the best-performing inequality-specific predictor configurations while maintaining the theoretical category structure. 

\section{Results}\label{sec4}
\subsection{Full Specification}\label{sec4-1}
\begin{figure}[h]
\centering
\includegraphics[width=\textwidth]{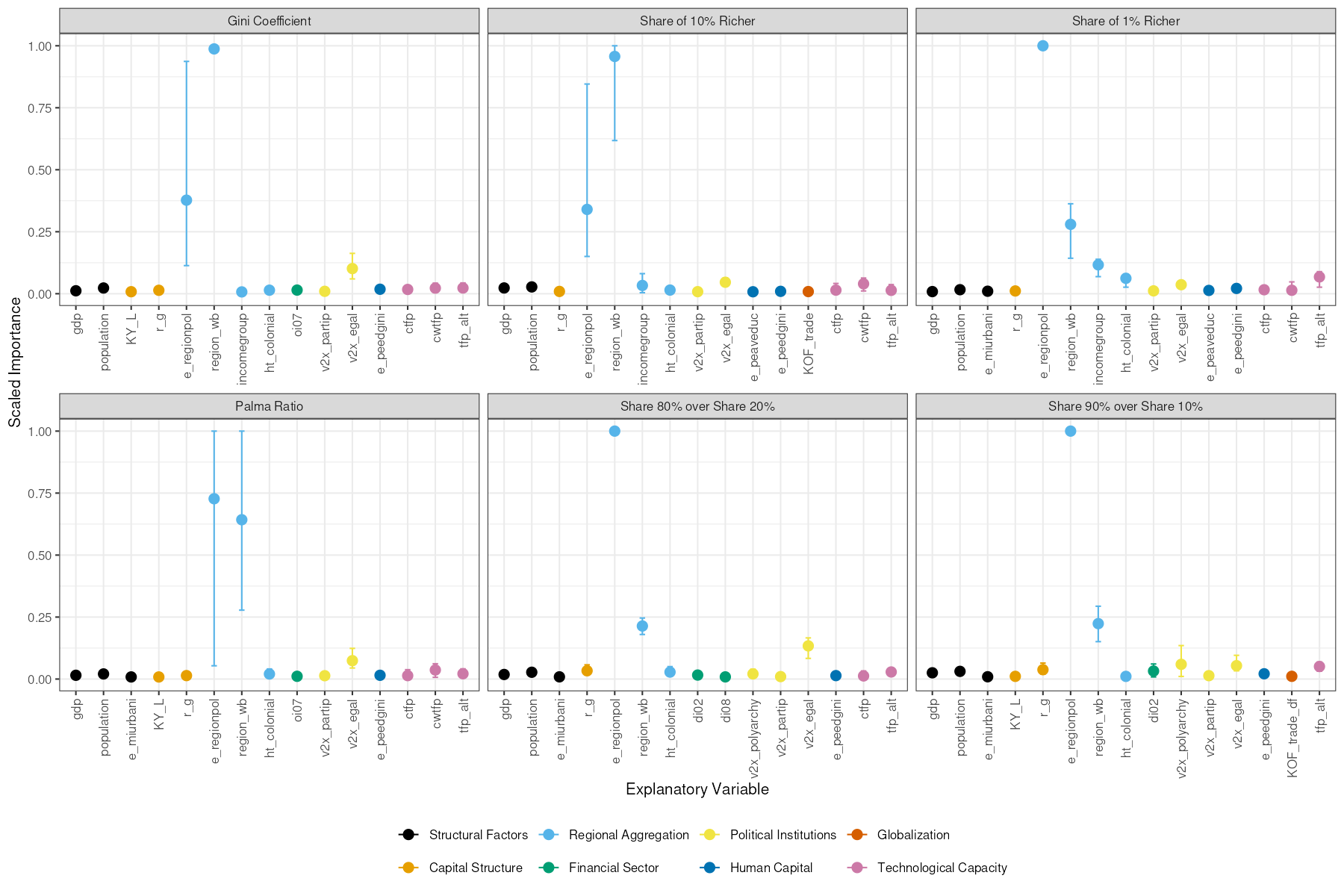}
\caption*{\footnotesize \textit{Notes:} The figure reports scaled feature importance from XGBoost models estimated using the full specification with all 35 candidate indicators. Only 10 indicators with the highest importance for each inequality measure are shown. Importance is normalized within each inequality measure to range between 0 and 1, preserving the relative ranking of indicators within each panel. Lines represent 95\% bootstrap confidence intervals. Each panel corresponds to a different inequality measure, and colors denote 8 conceptual categories.}
\caption{Scaled Feature Importance of the 10 Most Important Indicators Across Inequality Measures Under the Full Specification}
\label{fig:imps_all_sel}
\end{figure}

Figure \ref{fig:imps_all_sel} presents the results from the full model, showing the indicators with the highest feature importance for each inequality measure and providing an unrestricted assessment of their relative predictive importance based on \textit{XGBoost} feature importance scores. The corresponding results for all 35 candidate indicators are reported in Figure \ref{fig:imps_all}. Importance values are normalized within each inequality measure by rescaling them to a range between 0 and 1, allowing comparison of indicators within each panel while preserving their relative ranking in predicting that particular inequality measure.
 
The most striking result is the dominant predictive importance of a country’s regional classification across all inequality measures. In every case, either the politico-geographic classification ($e\_regionpol$) or the World Bank regional classification ($region\_wb$) is the most important predictor. 
This suggests that broad regional characteristics associated with geographic and historical context are strongly linked to cross-country differences in inequality, whereas other forms of aggregation considered in the analysis, such as income level or colonial history, display considerably lower predictive importance.

Beyond regional classification, indicators of political institutions also display relatively high predictive importance across several inequality measures, with the egalitarian component index ($v2x\_egal$) standing out most clearly for the Gini coefficient, the Palma ratio, and the S80/S20 ratio. Technology indicators also appear among more informative predictors, particularly for the top 1\% income share and the P90/P10 ratio. The other indicators are generally less important in the full specification, although the ratio of returns to capital to economic growth ($r\_g$) is relatively more predictive for the S80/S20 and P90/P10 ratios.

\subsection{Category-Representative Specification}\label{sec4-2}
To assess whether the predictive information contained in the full set of indicators can be retained in a more parsimonious specification, we construct a reduced model containing one representative indicator from each conceptual category. This also allows us to compare the predictive relevance of the different categories using a common specification in which each category is represented by a single indicator. We select the representative indicator from each category based on its average predictive importance across the six inequality measures, as estimated in the full model within each bootstrap iteration. We then select the indicators that are chosen most frequently across bootstrap iterations.

\begin{figure}[h]
\centering
\includegraphics[width=\textwidth]{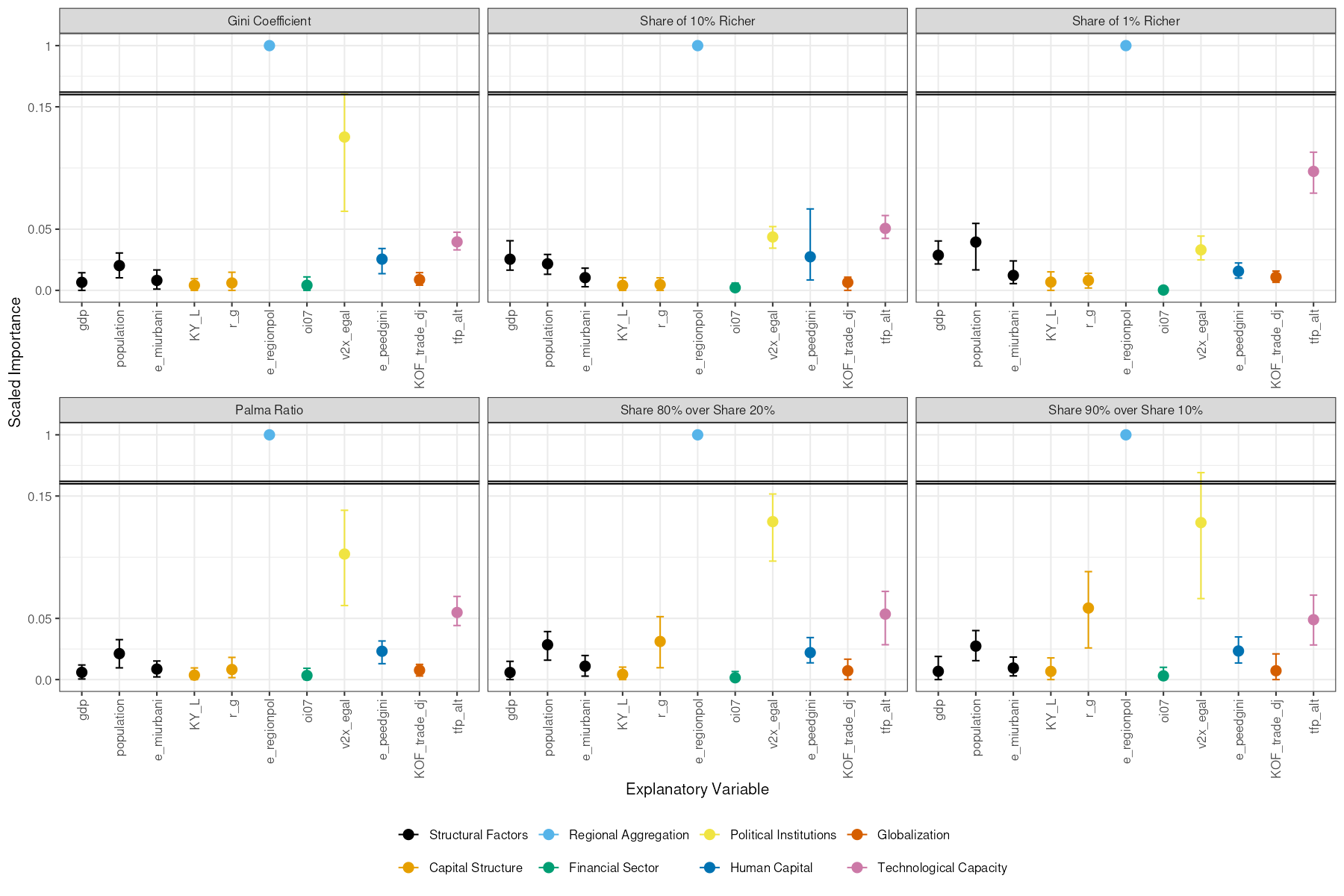}
\caption*{\footnotesize \textit{Notes:} The figure reports scaled feature importance from \textit{XGBoost} models estimated using the reduced specification. The model includes one representative indicator from each conceptual category, selected based on average predictive importance across the six inequality measures, together with the retained baseline controls. Importance is normalized within each inequality measure to range between 0 and 1, preserving the relative ranking of indicators within each panel. Each panel corresponds to a different inequality measure, and colors denote the conceptual categories. Lines represent 95\% bootstrap confidence intervals.}
\caption{Scaled Feature Importance using Category-Representative Predictors (Reduced Specification)}\label{fig:imps_sel_sel}
\end{figure}

Figure \ref{fig:imps_sel_sel} presents the results from the reduced specification, specifically, $di02$ as the representative measure of financial-sector development, $v2x\_egal$ for political institutions, $e\_peedgini$ for human capital, $KOF\_trade\_dj$ for globalization, and $tfp\_alt$ for technology. The model also includes the selected regional classification, indicators related to the structure of capital, and other retained indicators outside the six conceptual categories.
Results showing the indicators selected across all bootstrap iterations are presented in Figure \ref{fig:imps_sel}.

Consistent with the full specification, the politico-geographic regional classification ($e\_regionpol$) remains the dominant predictor across all inequality measures. It is followed by the egalitarian component index ($v2x\_egal$), which also consistently ranks among the most informative predictors. This suggests that political arrangements related to equality in political rights, resource distribution, and access to power are informative predictors of income inequality across the alternative measures considered.

The alternative TFP measure ($tfp\_alt$) is also among the more informative indicators, particularly for the top 1\% income share. This points to an association between differences in productivity and technology and upper-tail income concentration. The educational inequality indicator ($e\_peedgini$) has more modest predictive importance and shows relatively less variation across the inequality measures. The ratio of returns to capital to economic growth ($r\_g$) is relatively more predictive for the S80/S20 and P90/P10 ratios. Finally, the globalization indicator ($KOF\_trade\_dj$) shows moderate predictive importance across several inequality measures, whereas the selected financial-sector indicator remains comparatively less prominent in the reduced specification.

\subsection{Inequality Measure-Specific Specification}\label{sec4-3}
\begin{figure}[h]
\centering
\includegraphics[width=\textwidth]{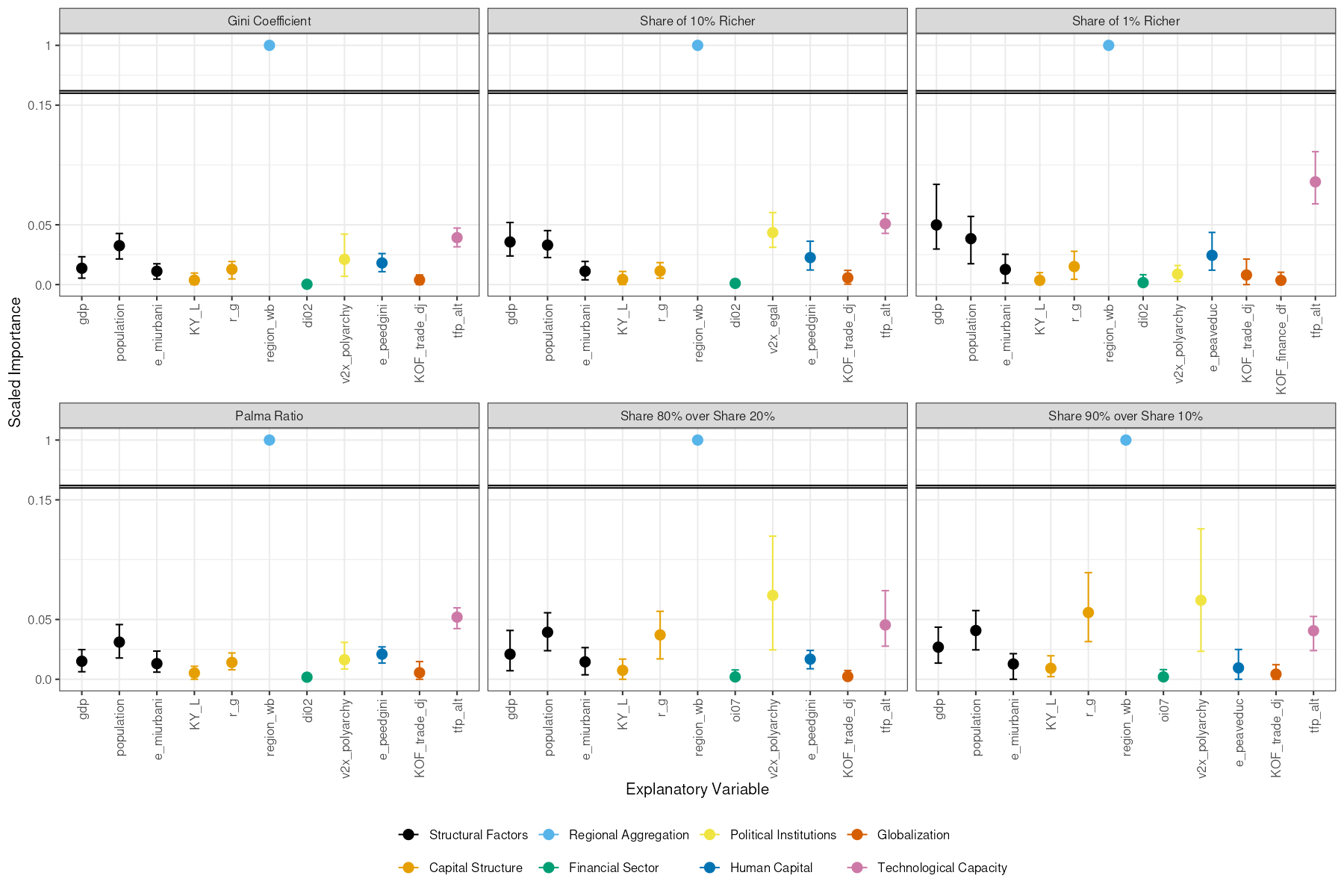}
\caption*{\footnotesize \textit{Notes:} The figure reports scaled feature importance from \textit{XGBoost} models estimated using the inequality measure-specific specification. For each inequality measure, the selected model is chosen from all possible combinations containing exactly one candidate indicator from each conceptual category, together with the retained baseline controls. 
Importance is normalized within each inequality measure to range between 0 and 1, preserving the relative ranking of indicators within each panel. Lines represent 95\% bootstrap confidence intervals. Each panel corresponds to a different inequality measure, and colors denote the conceptual categories.}
\caption{Scaled Feature Importance using Inequality-Specific Predictors (Inequality Measure-Specific Specification)}\label{fig:imps_bst_sel}
\end{figure}

The category-representative specification selects a common indicator from each conceptual category across all six inequality measures. The inequality measure-specific specification relaxes this restriction by allowing the selected indicator within each category to differ across inequality measures. This allows us to assess whether the patterns observed in the reduced model reflect broad average effects or measure-specific dynamics. For each inequality measure, we evaluate all possible combinations containing exactly one candidate indicator from each conceptual category, together with the retained indicators outside these categories. Figure \ref{fig:imps_bst_sel} presents the indicators selected most frequently across bootstrap iterations, while Figure \ref{fig:imps_bst} reports the results across all iterations.

Regional classification remains the dominant predictor under the inequality measure-specific specification, with the World Bank regional classification ($region\_wb$) selected for all six inequality measures. Political-institution and technology indicators also continue to rank among the most informative predictors, indicating that their importance is not driven solely by the use of a common set of indicators in the reduced specification. The selected political indicator varies across outcomes: $v2x\_polyarchy$ is selected for most measures, while $v2x\_egal$ is selected for the top 10\% income share. The alternative TFP measure ($tfp\_alt$) is particularly important for the top 1\% income share.

Human-capital indicators have more modest predictive importance and differ across measures, with educational inequality ($e\_peedgini$) selected more frequently than average years of education ($e\_peaveduc$). Financial-sector and globalization indicators remain comparatively less important. Overall, the specification reveals some measure-specific differences, while regional classification, political institutions, and technology remain the most consistently informative categories.

\subsection{Model Fit Across Specifications}\label{sec4-4}

Figure \ref{fig:preds} compares observed inequality values with the predictions from the three model specifications. All specifications display a very high degree of fit, with $R^2$ values above 0.98 for each inequality measure. The inequality measure-specific specification (“Best Within Measures”) has the highest $R^2$ in all six cases, with predictions that lie more closely along the 45-degree line. The improvement is most visible for the $P90/P10$ ratio and, to a lesser extent, the top 1\% income share and the $S80/S20$ ratio.

The category-representative specification (“Best Across Measures”) performs almost as well as the full specification despite including substantially fewer indicators. The differences in $R^2$ between these two specifications are very small across all inequality measures, indicating that the reduced model retains most of the predictive information contained in the full set of candidate indicators. Although allowing indicator selection to vary by inequality measure produces a further improvement in fit, the gains remain relatively modest. Overall, the comparison shows that a parsimonious set of category-representative indicators can provide predictions that are close to those obtained from both the full and inequality-measure-specific specifications.

\section{Discussion}\label{sec12}
The results show both consistency and important differences in the predictors of alternative measures of income inequality. Across all model specifications and inequality measures, regional aggregation remains the single most important predictor of future inequality outcomes, with the politico-geographic classification playing a particularly prominent role. This finding is consistent with comparative evidence showing that countries form distinct and persistent regional patterns of inequality \citep{MannRiley2007,Andreescu2025}. Similar differences are also visible descriptively in our data, as Figure \ref{fig:ineq_region} shows clear variation in income inequality across politico-geographic regions. However, because regional aggregation combines a wide range of political, institutional, economic, historical, and geographical characteristics, the analysis provides only limited insight into which region-specific factors account for its predictive importance. To test the importance of these other unobserved regional factors, we re-estimated otherwise identical \textit{XGBoost} models after removing the regional classification indicators. Excluding region increases hold-out MSE and lowers hold-out $R^2$ for all six inequality measures, indicating that regional classification contributes important predictive information beyond the other indicators included in the analysis (Figure \ref{fig:mse_rsq_tests_1}).

At the same time, among the political institution indicators, the egalitarian component index ($v2x\_egal$) consistently ranks among the most informative predictors across inequality measures, particularly for the Gini coefficient, the Palma ratio, and the S80/S20 ratio. This suggests that equality in political rights, access to power, and resource distribution across social groups is closely associated with broader patterns of income inequality, consistent with research emphasizing the role of political institutions in shaping economic opportunities and the distribution of economic gains \citep{Acemoglu2001,Acemoglu2005,Acemoglu2012}.

Technology also provides important predictive information, with the total factor productivity measure derived from the growth of capital services ($tfp\_alt$) standing out particularly for the top 1\% income share. This suggests that differences in productivity and technological capacity are especially closely associated with income concentration at the very top. This is consistent with findings from labour economics literature showing that technological change can alter the demand for skills and widen earnings differences across workers and occupations \citep{Krueger1993,Acemoglu2002}.

Human capital has modest and relatively stable predictive importance across inequality measures, although educational inequality ($e\_peedgini$) appears somewhat less informative for the top 1\% income share than for the other measures. Its selection as the representative human capital indicator among others suggests that how education is distributed may be more informative for predicting inequality than the national average alone. The literature shows that countries with higher average levels of educational attainment tend to display lower levels of inequality \citep{Gregorio2002}, but at the same time, the effects of educational expansion depend on labour-market institutions, the composition of labour demand, and broader economic conditions \citep{Katz1992,Autor2014}. Therefore, increases in education may not automatically reduce inequality when productivity gains and labour-market rewards remain concentrated in particular sectors or occupations \citep{Makhlouf2023,Chang2024}.

Among the capital structure indicators, the ratio of returns to capital to economic growth ($r\_g$) is relatively more informative for the S80/S20 and P90/P10 ratios, consistent with Piketty’s argument that high returns to capital can widen income gaps \citep{Piketty2014}.

The results also provide a more nuanced picture of financial-sector development and globalization. Although both are widely discussed in the inequality literature, their predictive importance is less consistent than that of the other indicators. The selected financial-sector indicator ($di02$) remains comparatively weak across inequality measures, whereas trade globalization ($KOF\_trade\_dj$) shows moderate importance for several outcomes. Financial development may affect inequality through access to credit and financial assets \citep{Greenwood1990,deHaan2017}, while globalization may influence wages and employment through trade and import competition \citep{Helpman2010, Autor2013}.

Overall, the findings show that the predictors of income inequality are neither entirely common across measures nor completely measure-specific. Regional aggregation, political institutions, and technology provide the most consistent predictive information, while the relevance of human capital, returns to capital, financial-sector development, and globalization varies more across inequality measures. The results therefore highlight the importance of considering multiple measures of inequality, as relying on a single measure may fail to capture differences in the factors associated with different parts of the income distribution.

These findings should nevertheless be interpreted in light of several limitations. First, the analysis is predictive rather than causal, so the identified indicators should not be understood as direct causes of inequality. Second, although regional classification is highly predictive, it is broad and does not reveal which underlying regional characteristics drive its predictive importance. 
Third, because the analysis uses country-level data, it cannot fully capture within-country mechanisms that generate inequality at the household or individual level.
Finally, although we selected a broad range of indicators from the literature, they still do not capture all the factors that may affect inequality. 


\section{Conclusion}\label{sec13}
This study examines the predictive importance of 35 indicators across six measures of income inequality using \textit{XGBoost} models for 88 countries from 1970 to 2017. We find that politico-geographic regional classification is the strongest predictor across all specifications. Political institutions and technology provide the most consistent additional information, while human capital plays a more modest but relatively stable role. Returns to capital are more informative for the S80/S20 and P90/P10 ratios, whereas financial-sector development and globalization are generally less prominent.

The findings show that the importance of different indicators depends partly on how inequality is measured. They also show that a relatively small set of carefully selected indicators retains most of the predictive information contained in the full specification, allowing for a more focused and parsimonious analysis of cross-country inequality.

From a practical perspective, the results suggest that informative predictions of income inequality can be obtained without relying on a large and complex set of indicators. The approach can also support inequality monitoring by identifying which indicators are most informative for different dimensions of the income distribution.

\backmatter

\bmhead{Supplementary information}



\bmhead{Acknowledgements}

\section*{Declarations}
\begin{itemize}
\item \textbf{Funding:}This research received no external funding.
\item \textbf{Conflict of interest/Competing interests:}The authors declare that they have no financial or non-financial competing interests that are directly or indirectly related to this work.
\item \textbf{Ethics approval and consent to participate:} Not applicable.
\item \textbf{Consent for publication:} Not applicable.
\item \textbf{Data availability:}
\item \textbf{Materials availability:}Not applicable.
\item \textbf{Code availability:}
\item \textbf{Author contribution:}
\end{itemize}

\noindent








\bibliography{sn-bibliography}

\appendix

\renewcommand{\thefigure}{A\arabic{figure}}
\setcounter{figure}{0} 

\begin{figure}[h]
\centering
\includegraphics[width=\textwidth]{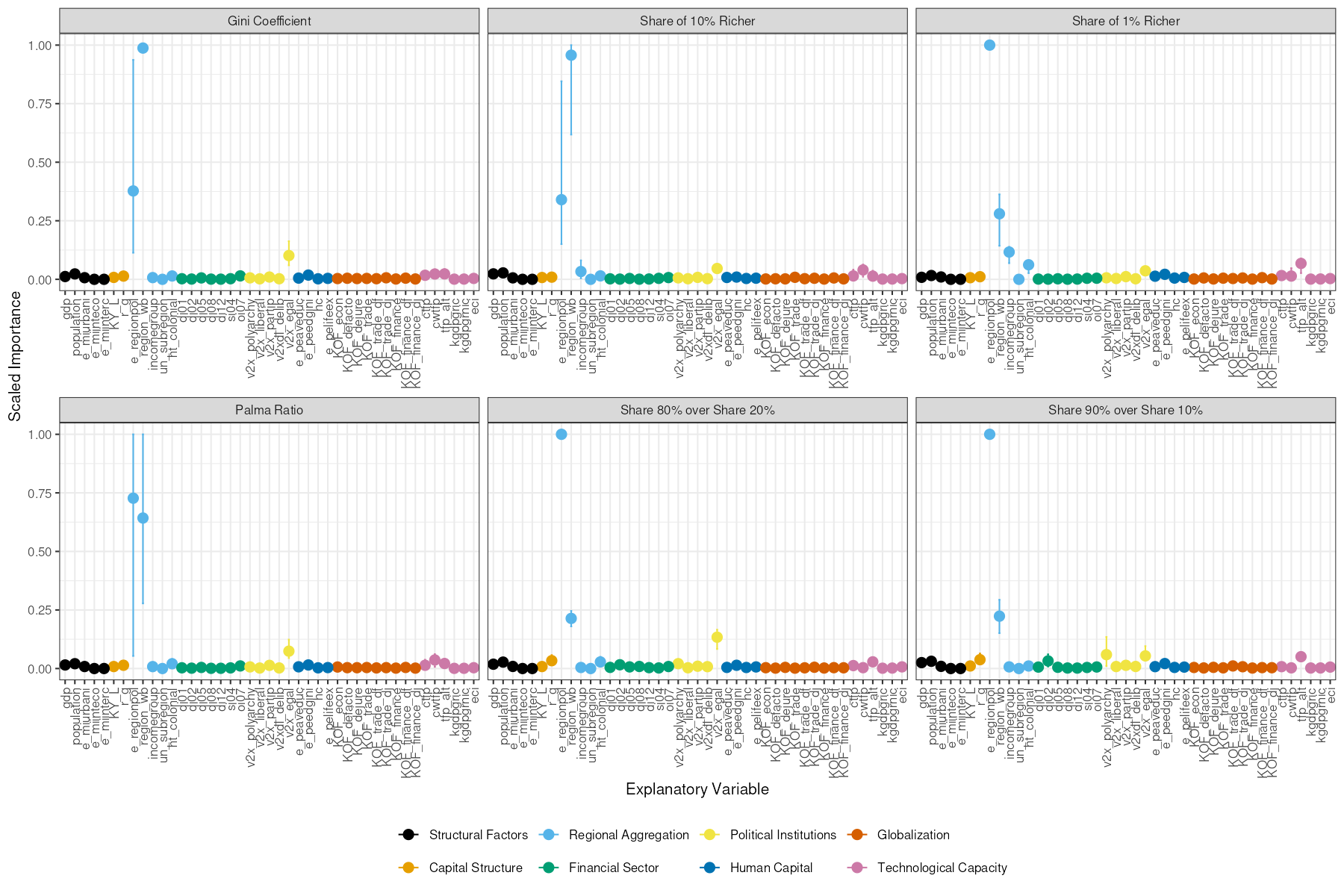}
\caption*{\footnotesize \textit{Notes:} The figure reports scaled feature importance from XGBoost models estimated with the full specification, including all 35 candidate indicators. Importance is normalized within each inequality measure to range between 0 and 1, preserving the relative ranking of indicators within each panel. Each panel corresponds to a different inequality measure, and colors denote the conceptual categories.}
\caption{Scaled Feature Importance Across Inequality Measures Under the Full Specification - All Variables Across All Bootstrap Iterations}
\label{fig:imps_all}
\end{figure}

\begin{figure}[h]
\centering
\includegraphics[width=\textwidth]{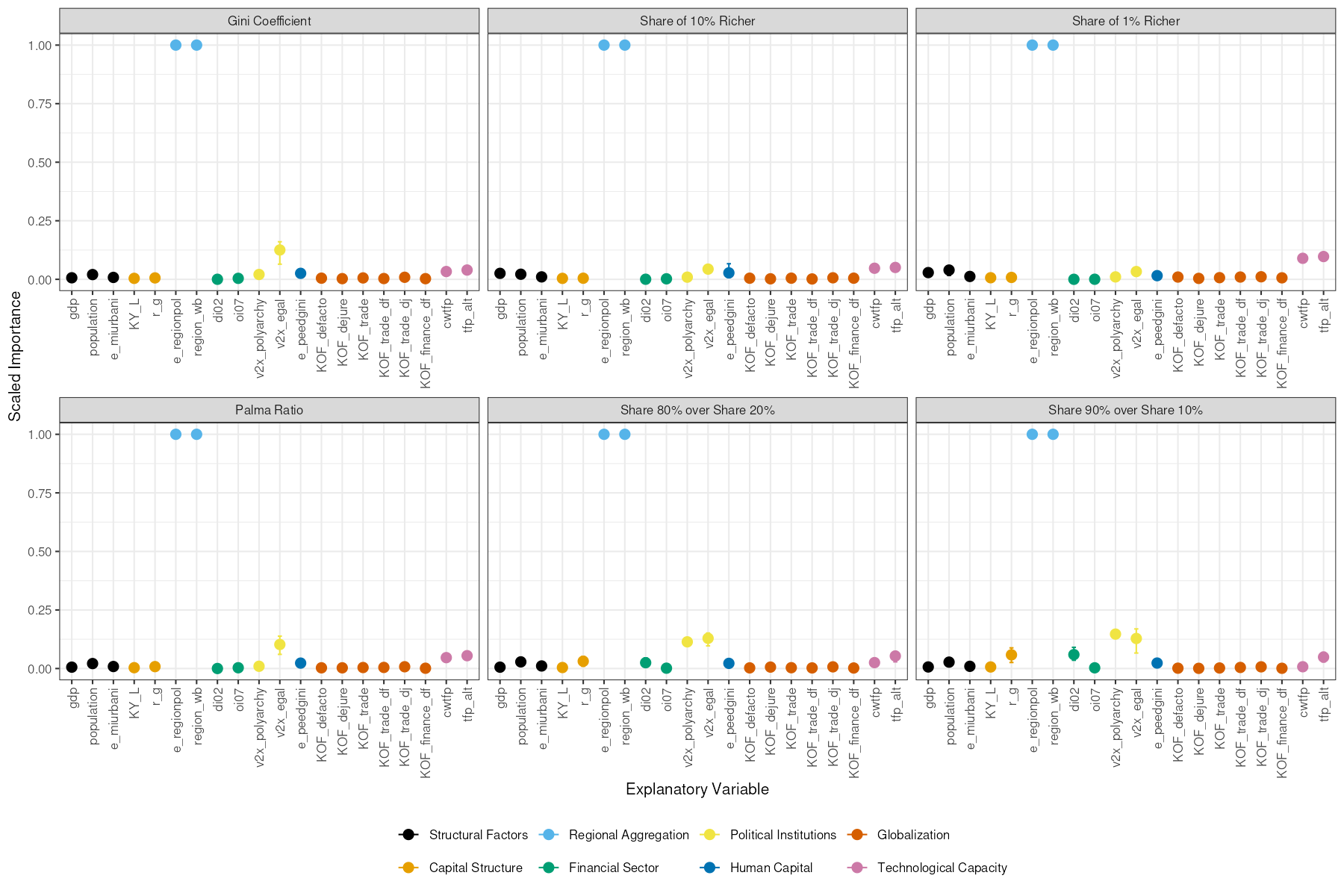}
\caption*{\footnotesize \textit{Notes:} The figure reports scaled feature importance from \textit{XGBoost} models estimated using the reduced specification. The model includes one representative indicator from each conceptual category, selected based on average predictive importance across the six inequality measures, together with the retained baseline controls. Importance is normalized within each inequality measure to range between 0 and 1, preserving the relative ranking of indicators within each panel. Each panel corresponds to a different inequality measure, and colors denote the conceptual categories.}
\caption{Scaled Feature Importance Using Category-Representative Predictors (Reduced Specification - All Selected Variables Across All Bootstrap Iterations)}
\label{fig:imps_sel}
\end{figure}

\begin{figure}[h]
\centering
\includegraphics[width=\textwidth]{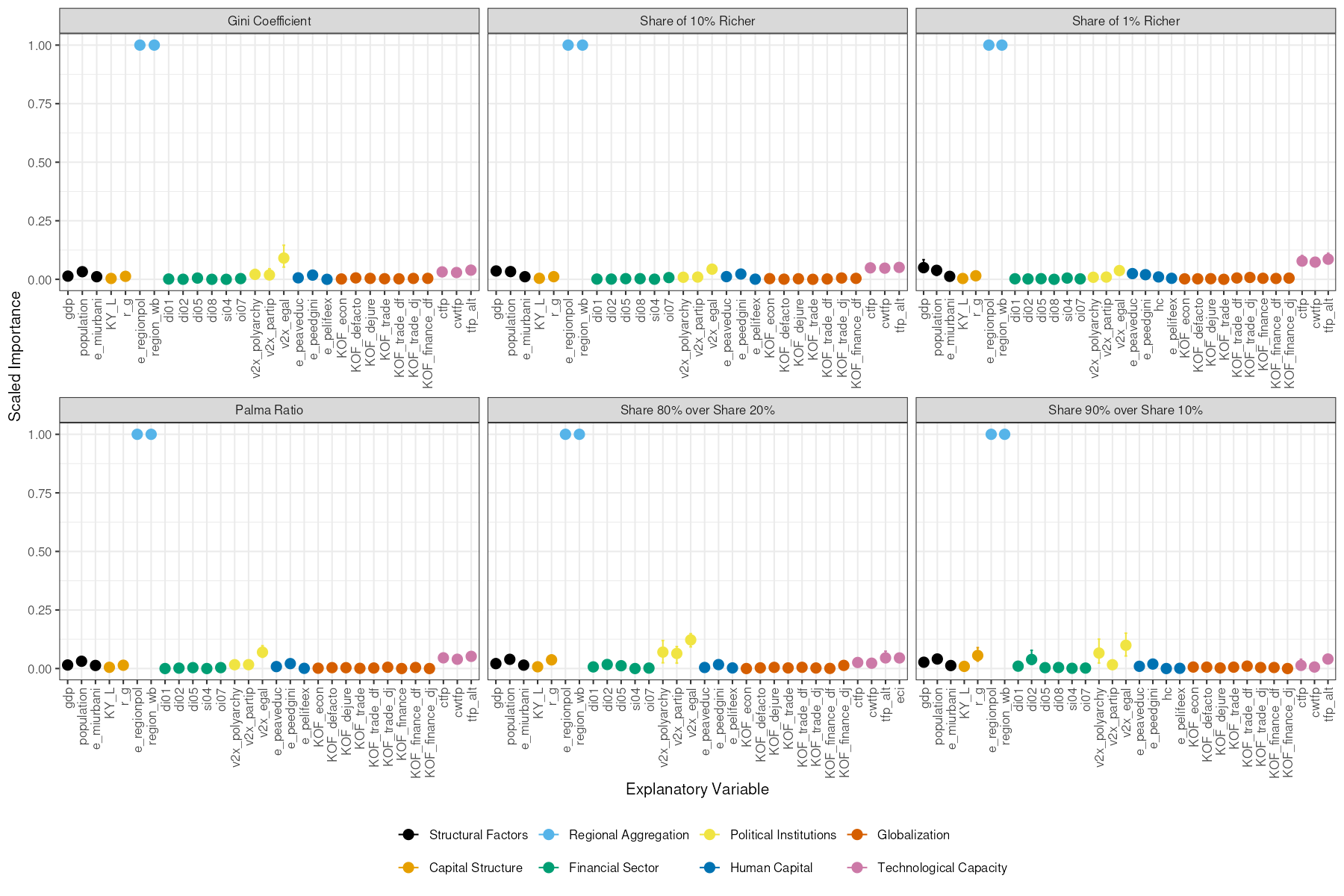}
\caption*{\footnotesize \textit{Notes:} The figure reports scaled feature importance from \textit{XGBoost} models selected under the inequality-specific specification. For each inequality measure, the model is chosen from all combinations containing exactly one candidate variable from each conceptual category, together with the retained general controls. The selected model minimizes the combined mean squared prediction error across the training and hold-out samples. Importance is normalized within each inequality measure (0–1). Colors denote thematic variable categories.}
\caption{Scaled Feature Importance using Inequality-Specific Predictors (Inequality Measure-Specific Specification - All Selected Variables Across All Bootstrap Iterations)}
\label{fig:imps_bst}
\end{figure}

\begin{figure}[h]
\includegraphics[width=\textwidth]{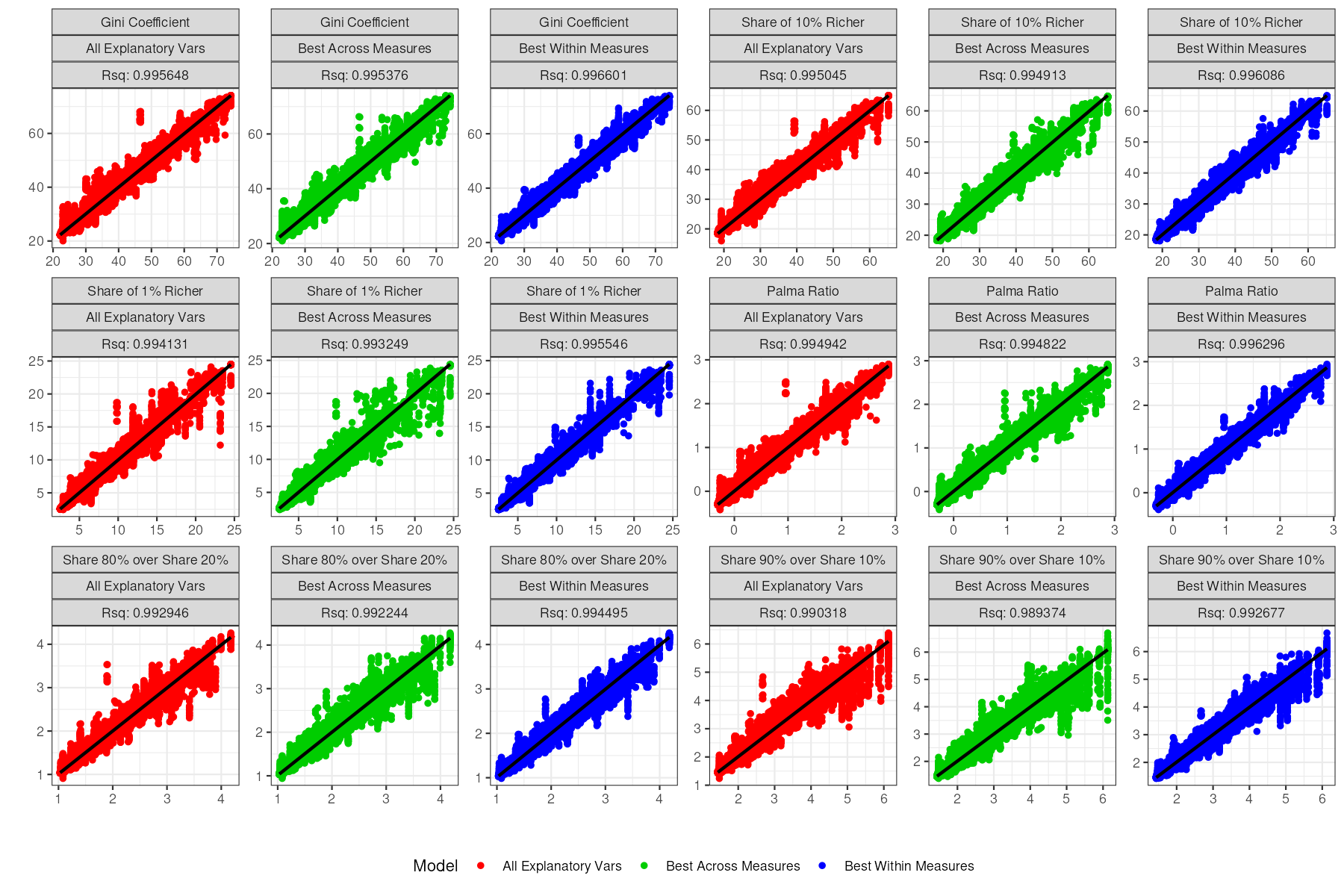}
\caption*{\footnotesize \textit{Notes:} The figure plots predicted values against observed inequality values for each of three model specifications and six inequality measures. Predictions are obtained from the fitted \textit{XGBoost} models under three specifications. “All Explanatory Variables” refers to the full specification, “Best Across Measures” to the reduced specification using category-representative predictors, and “Best Within Measures” to the inequality-specific specification. The solid line represents the 45-degree line corresponding to perfect prediction.}
\caption{Predicted versus Observed Inequality across Model Specifications}\label{fig:preds}
\end{figure}

\begin{figure}[h]
\centering
\includegraphics[width=\textwidth]{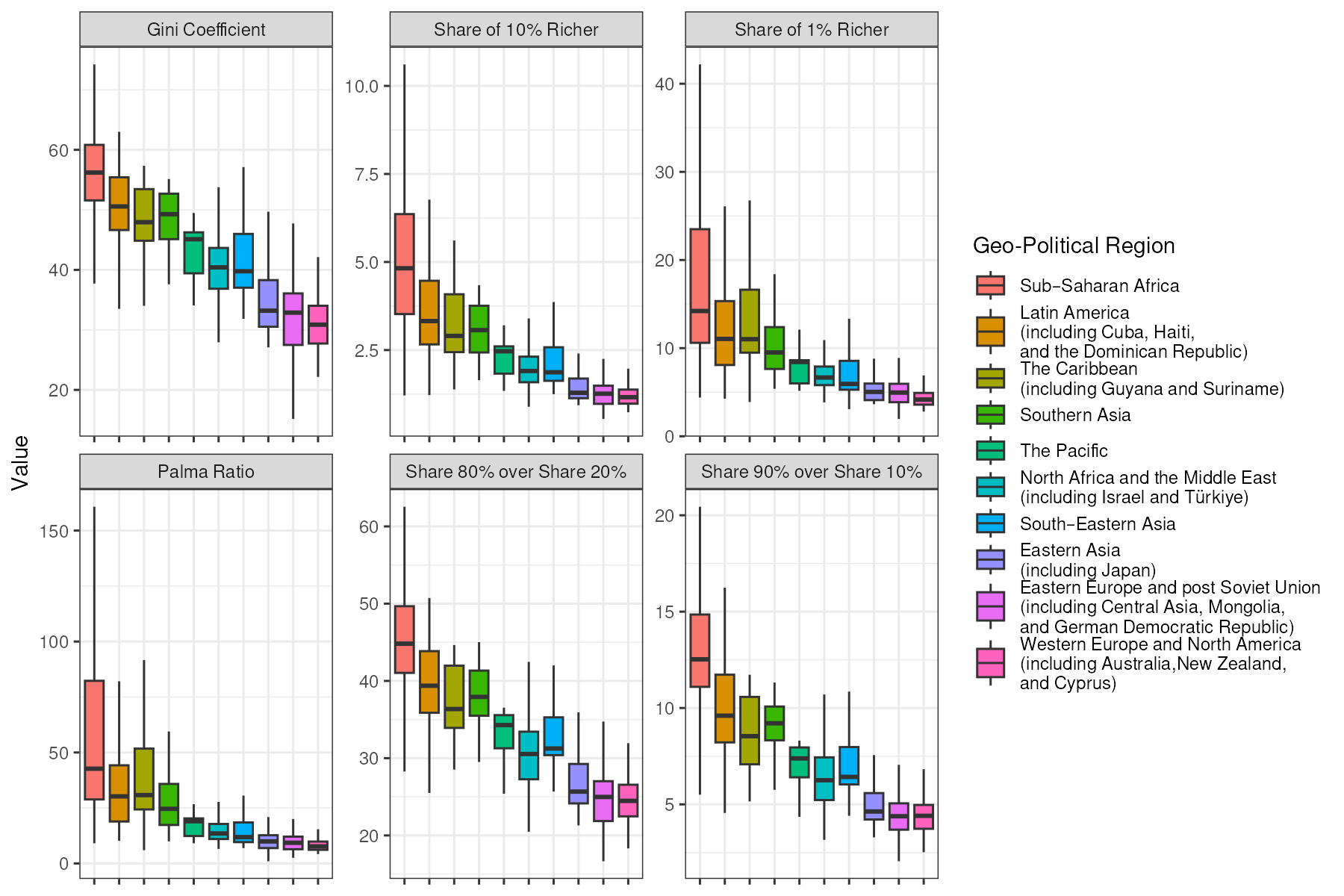}
\caption*{\footnotesize \textit{Notes:} The figure shows the distribution of six income inequality measures across the ten politico-geographic regions defined by the V-Dem regional classification ($e\_regionpol$). Boxes indicate the interquartile range, horizontal lines denote the median, and whiskers show the remaining distribution excluding outliers. Each panel corresponds to a different inequality measure. The figure is descriptive and does not imply a causal relationship between regional classification and income inequality.}
\caption{Distribution of Income Inequality Across Politico-Geographic Regions for Different Inequality Measures}\label{fig:ineq_region}
\end{figure}


\begin{figure}[h]
\centering
\makebox[\textwidth][c]{%
  \includegraphics[width=1.2\textwidth]{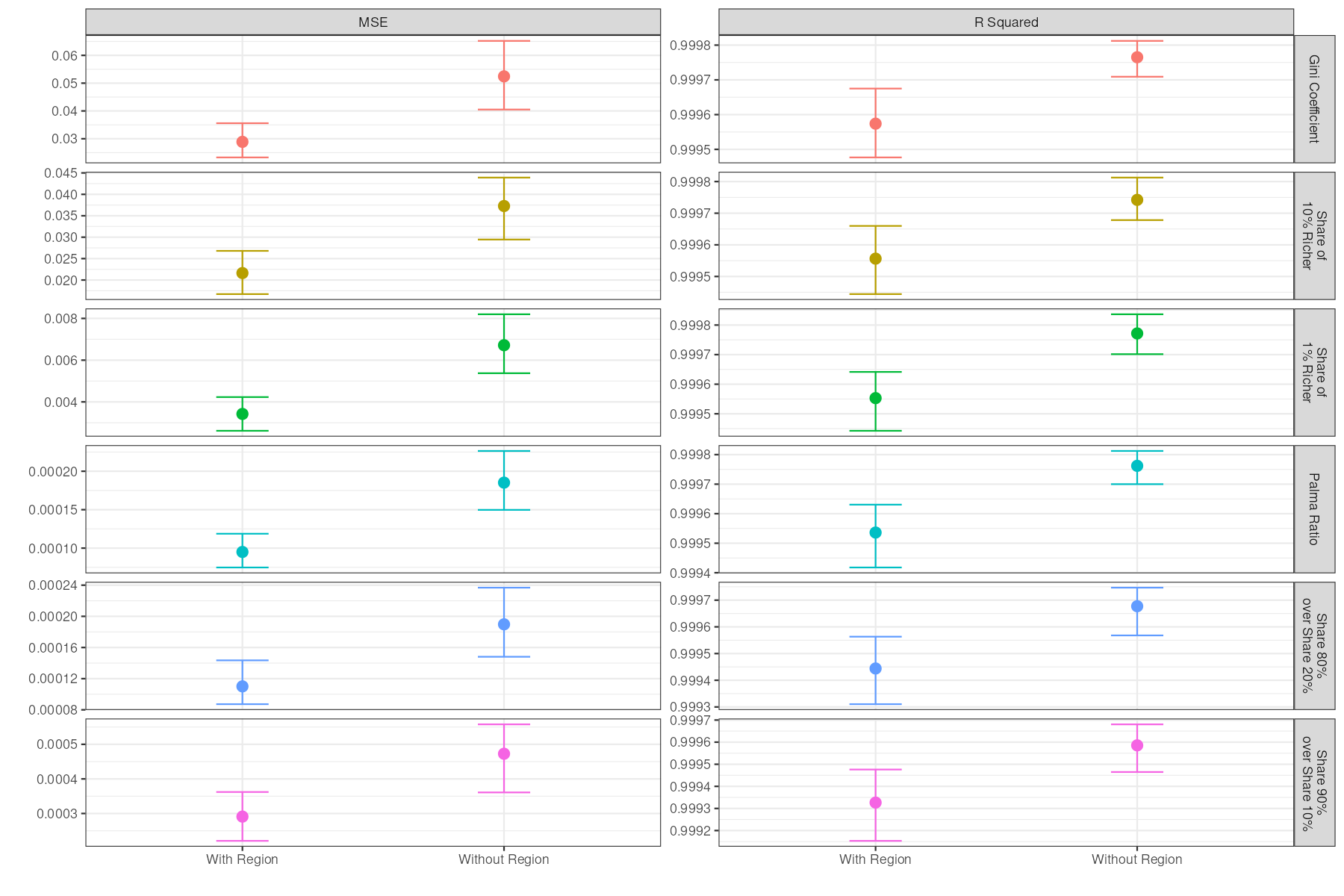}}
\caption*{\footnotesize \textit{Notes:}The figure compares hold-out MSE and $R^2$ from otherwise identical \textit{XGBoost} models estimated with and without the regional classification indicators. Hold-out MSE is the mean squared prediction error calculated using the 30\% of observations not used to train the model; lower values indicate better predictive performance. Higher hold-out $R^2$ values indicate a closer correspondence between observed and predicted inequality. Points report average performance across bootstrap iterations, and lines represent 95\% confidence intervals. The same training–hold-out split is used for both models within each iteration.}
\caption{Predictive Performance of XGBoost Models With and Without Regional Classification Across Inequality Measures}\label{fig:mse_rsq_tests_1}
\end{figure}

\end{document}